\documentclass[onecolumn,aps,nofootinbib,superscriptaddress]{revtex4-2}
\usepackage{graphicx}
\usepackage{xcolor}
\usepackage{soul}
\usepackage{url}

\usepackage{breakurl}
\usepackage[breaklinks]{hyperref}
\usepackage{multirow}
\usepackage{setspace}
\usepackage{amsmath, amssymb, amsthm}
\usepackage{cleveref}
\usepackage{physics}
\usepackage{booktabs}
\usepackage{footnote}
\usepackage{footmisc}
\usepackage{longtable}

\begin{document}

\title{Unmasking the mask studies: why the effectiveness of surgical masks in preventing respiratory infections has been underestimated}
\author{Pratyush K. Kollepara}
\thanks{These two authors contributed equally.}
\affiliation{New England Complex Systems Institute, Cambridge, MA}
\affiliation{Department of Physics, BITS Pilani K K Birla Goa Campus, Goa, India}
\author{Alexander F. Siegenfeld}
\thanks{These two authors contributed equally.}
\affiliation{New England Complex Systems Institute, Cambridge, MA}
\affiliation{Department of Physics, Massachusetts Institute of Technology, Cambridge, MA}
\author{Nassim Nicholas Taleb}
\affiliation{Tandon School of Engineering, New York University, Brooklyn, NY}
\author{Yaneer Bar-Yam}
\affiliation{New England Complex Systems Institute, Cambridge, MA}

\begin{abstract}
Face masks have been widely used as a protective measure against COVID-19. However, pre-pandemic empirical studies have produced mixed statistical results on the effectiveness of masks against respiratory viruses.  The implications of the studies' recognized limitations have not been quantitatively and statistically analyzed, leading to confusion regarding the effectiveness of masks.  Such confusion may have contributed to organizations such as the WHO and CDC initially not recommending that the general public wear masks.  Here we show that when the adherence to mask-usage guidelines is taken into account, the empirical evidence indicates that masks prevent disease transmission: all studies we analyzed that did not find surgical masks to be effective were under-powered to such an extent that even if masks were 100\% effective, the studies in question would still have been unlikely to find a statistically significant effect. We also provide a framework for understanding the effect of masks on the probability of infection for single and repeated exposures.  The framework demonstrates that more frequently wearing a mask provides super-linearly compounding protection, as does both the susceptible and infected individual wearing a mask.  This work shows (1) that both theoretical and empirical evidence is consistent with masks protecting against respiratory infections and (2) that nonlinear effects and statistical considerations regarding the percentage of exposures for which masks are worn must be taken into account when designing empirical studies and interpreting their results.
\end{abstract}

\maketitle

In 1910, one of the first western-trained Chinese physicians adapted surgical masks for use against a respiratory plague that killed more than 60,000 people in four months~\cite{goodman_2020}. The logic behind their function is transparent: a mask can block some viral or bacterial particles from entering and/or dispersing from the wearer's respiratory tract. They have been used for prevention in a wide range of disease outbreaks and medical settings, and there is currently a general consensus that surgical and cloth masks help prevent infected individuals from spreading COVID-19~\cite{CDC_mask, WHO_mask}. Surprisingly, given the logic of their utility, there is less of a consensus that surgical/cloth masks also protect the wearer and many government health organizations did not initially recommend wearing them during the early months of the COVID-19 pandemic~\cite{FT2020, mask_use_gov}.   

\begin{figure}[h]
\centering
\includegraphics[scale=1, trim={0 0 0 0}, clip]{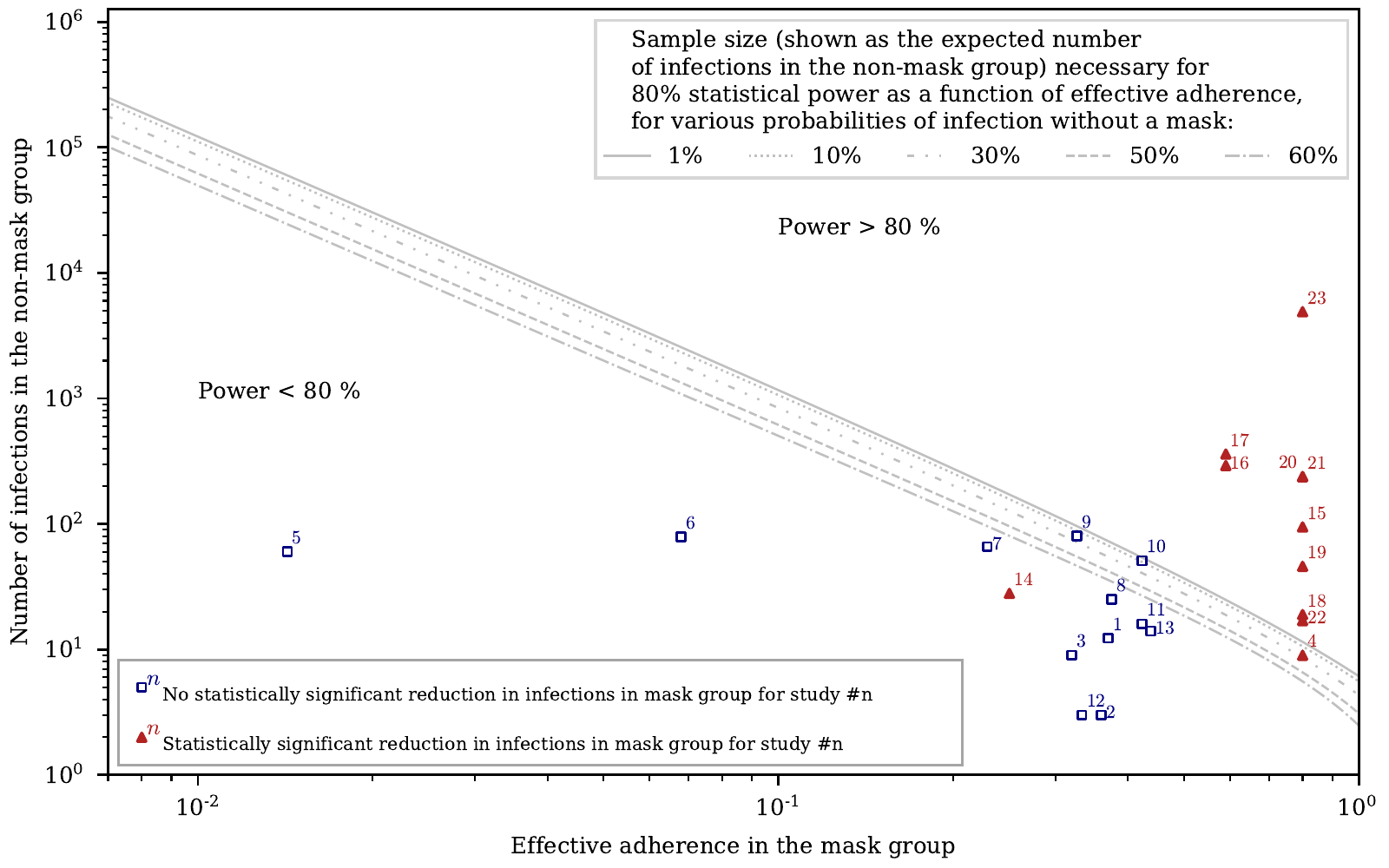}
\caption{The effective adherence and sample sizes of studies that found masks to be effective (red triangles) and those that did not (blue squares). Empirical studies with higher levels of statistical power consistently show that masks protect the wearer; studies with lower statistical power are mixed, as would be expected. The statistical power depends on the sample size, the effective adherence (i.e. mask effectiveness multiplied by the fraction of exposures for which masks are on average worn in the mask group), and probability of being infected without a mask in the setting of the study.  Each curve depicts the required sample size (expressed as the expected number of infections in the non-mask group) as a function of effective adherence in order for the study to have a power of 80\%, i.e. an 80\%  probability of finding a statistically significant result ($p<0.05$, two-tailed).  The curves are calculated assuming equal-sized non-mask and mask groups, with each curve representing a different probability of infection in the non-mask group; the total sample size is thus the expected number of infected individuals in the non-mask group (which is what is plotted) divided by the probability of infection and then multiplied by two.  We assume that the probability of infection is reduced linearly with increasing effective adherence (see \cref{fig:p_inf}); this assumption will underestimate the true necessary sample size.   The scattered data points depict the size and effective adherence of studies taken from a recent systematic review~\cite{Brainard2020}; the study numbers correspond to those in the first column of \cref{fig:power_table}.  The effective adherence for the studies are overestimated by assuming that masks are 100\% effective; even with this assumption, the studies numbered 1 through 14 were found to have less than 80\% statistical power.  Note that the power of the studies (reported in \cref{fig:power_table}) depend on the probability of infection in the non-mask group and the size of the mask group in addition to the information present in this figure; the location of the studies in this figure relative to the curves should therefore be considered only approximately.  In particular, some of the studies have smaller overall sample sizes than implied by this figure, due to their non-mask and mask groups not being equal in size. Mathematical details can be found in the Appendix.}
\label{fig:power_plot}
\end{figure}

It is well established that surgical and cloth masks partially block virus-containing airborne droplets of various sizes~\cite{Booth2013, gawn2008, lindsley2012, Milton2013, Alsved2020, vanderSande2008, Leung2020, Davies2013, Konda2020, Chughtai2020}. Cloth masks, surgical masks, respirator masks (e.g. N95), and powered air-purifying respirators are understood to be capable of providing increasing levels of protection.  The amount of virus transmitted between an infected and a susceptible individual is therefore expected to be reduced if either is wearing a mask, with both wearing masks giving the best protection. However, this straightforward inference has been difficult to establish in experimental studies. 
Here we analyze why some experimental studies find masks to be effective while others do not.  We determined that the studies that did not find surgical masks to be effective were under-powered to such an extent that even if the masks were 100\% effective, they still would have been unlikely to find a statistically significant result.  Statistical power is the probability that a study will find a statistically significant result if its intervention does in fact have a certain effect.   Our results concerning the statistical power of mask studies are summarized in \cref{fig:power_plot}, which shows that all studies that had a large enough sample size and/or adherence for 80\% power (above and to the right of the gray lines) show a statistically significant reduction in infections among mask-wearers.  As would be expected, most studies with less statistical power (towards the lower left) did not find a statistically significant effect. 
We also provide a framework for understanding the nonlinear effects of mask-wearing on the probability of infection.  Experiments that do not take such factors into account provide misleading results unless interpreted carefully.  While the precautionary principle~\cite{Mask_review, Howard2020} would recommend the use of masks during the COVID-19 pandemic in any case (due to the asymmetric risks of using vs. not using masks), the analyses we provide gives consistency to theoretical analyses, experimental studies, and epidemiological recommendations. 

\section*{Statistical power}

Some empirical studies find masks to be effective in preventing disease transmission while others do not~\cite{Mask_review, MacIntyre2009, COWLING2010, Aledort2007, Xiao2020, Brainard2020, Chu2020}.  However, due to poor statistics, even the studies with negative results are not inconsistent with masks being highly effective.  While some of the studies conducted a power analysis to estimate the sample size required to obtain a statistically significant result with 80\% probability (i.e. to achieve 80\% power, the standard level by convention), these power analyses did not take in to account the possibility of low adherence (i.e. masks being worn for a low percentage of exposure events) and/or the possibility of a very low probability of infection even without a mask.  When we consider such factors, none of the studies we analyze that did not find masks to be effective had sufficient statistical power. 

The sample of studies we consider is taken from a recent systematic review~\cite{Brainard2020}; see~\cref{tab:exc} for a list of studies that were excluded and why.  Most of the studies we examine measure whether surgical masks protect the wearer; the exceptions are studies nos. 8, 12 and 15, which measure whether masks prevent the wearer from infecting others, and studies nos. 1, 7, 13, and 18, in which both the susceptible and infected individuals sometimes wore masks (see \cref{tab:adh}).

\begin{figure}
\centering
\includegraphics[scale=0.95, trim={0 0cm 0 0cm}, clip]{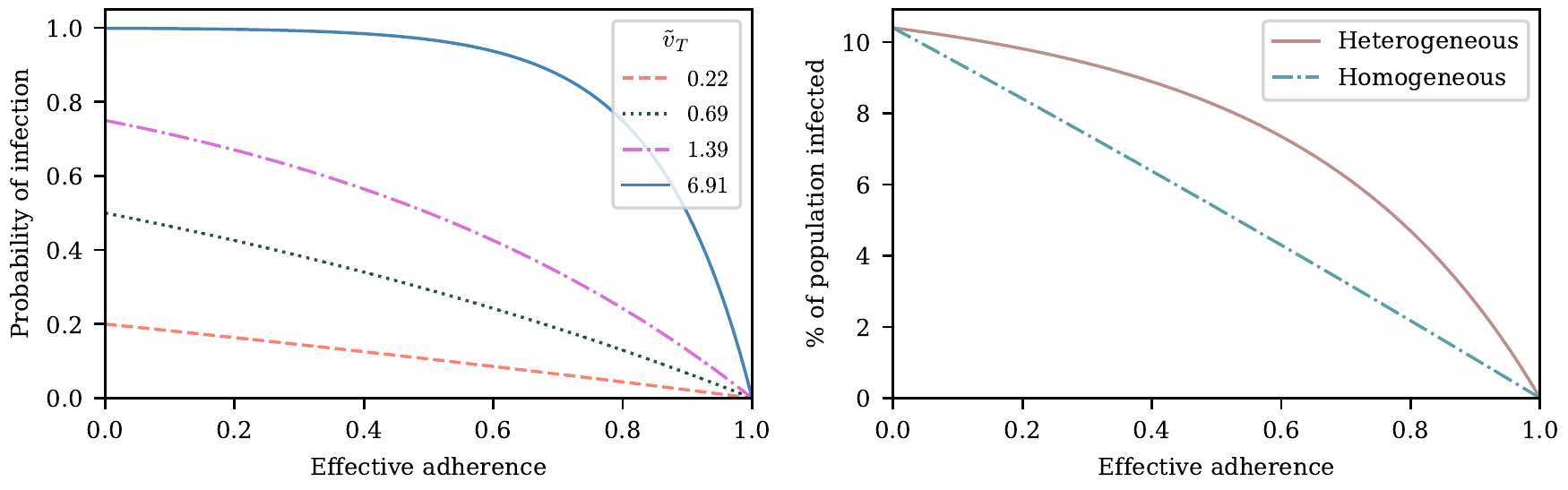}
\caption{\textbf{Left:} A susceptible individual's probability of infection as a function of effective adherence $\alpha\gamma$ (mask effectiveness $\gamma$ multiplied by the fraction of exposures for which the mask is worn $\alpha$) for various values of that individual's total effective exposure $\tilde v_T$ (the total effective exposure is proportional to the number of exposure events).  Note that if a mask were 100\% effective ($\gamma=1$), then the effective adherence would simply equal the adherence $\alpha$. The curves denoting the infection probabilities are given by $1 - e^{-(1-\alpha \gamma)\tilde v_T}$ (\cref{eq:onemask}).  For high values of $\tilde v_T$, the infection probability is nonlinear in the adherence, while for low values of $\tilde{v}_T$, the infection probability decreases approximately linearly with adherence. 
\textbf{Right:}  For a group of individuals (e.g. in an arm of a study), the total effective exposure will in general vary from individual to individual such that even if on average the total effective exposure is relatively low, it may be high for the individuals who make up the bulk of those being infected.  Thus, while using the average effective exposure would predict an approximately linear decrease of infection probability with increasing adherence, such an approach may overestimate the expected effect of partial mask usage.  The dashed curve depicts the expected percentage of infected individuals for the homogeneous case in which everyone experiences the same total effective exposure, whereas the solid curve depicts a case in which the exposure is heterogeneous; in both cases, the percentages of individuals that would be infected without masks (e.g. in a control group) are identical (approximately 10\%).}
\label{fig:p_inf}
\end{figure}

In order to account for adherence, we make two conservative assumptions that will result in our overestimating the studies' statistical powers.  First, we assume that the degree to which a mask reduces the probability of infection is proportional to the fraction of exposures for which it is worn (e.g. we assume wearing a mask half as often provides half as much protection); in fact, wearing a mask half as often will reduce the probability of infection by less than half as much (see \cref{fig:p_inf}), meaning that we overestimate the statistical power of these studies.  Second, the numbers we calculate represent the power the studies would have had were masks 100\% effective (i.e. were it impossible to become infected while wearing a mask).  To the extent that masks are less than 100\% effective, even larger sample sizes would be needed.

For example, a randomized control trial (RCT) at the Hajj pilgrimage~\cite{Alfelali2020} assumed a reduction in infection rate from $12\%$ to $6\%$ in order to determine the sample size necessary for a statistical power of $80 \%$.  After taking into account that the randomization was done by cluster (i.e. tent) rather than individual, the required sample size was $\sim 6000$.  However, the study reports that individuals in the intervention group on average wore masks for far less than half the time.  Under these conditions, even with perfectly effective masks, a 50\% reduction in infection probability is impossible.   Adherence can be estimated by the product of the fraction of people who wore masks and the fraction of time during which exposure is possible for which masks were used.  The data reported in the study indicate an adherence in the mask group of $3.2\%$, which could cause at most a $3.2\%$ reduction in the probability of infection.  However, the adherence in the control group was 1.8\%, meaning that the maximum possible expected reduction in infection between the two groups would be $\frac{0.032-0.018}{1-0.018}=0.014$ (\cref{eq:net_adh}).  Thus the effective adherence value used for this study is 0.014.  In addition, the probability of infection without masks is reported to be quite low ($2 \%$). Under these conditions, the required sample size to achieve the desired statistical power of 80\% would be $\sim$ 7.8 million (with individual randomization; with cluster randomization an even larger number of participants is needed).  Power analyses for other studies and the methods used are described in the appendix and summarized in \cref{fig:power_plot,fig:power_table,tab:adh}. 

Other factors such as false positives may also limit statistical power.  For instance, a recent study~\cite{Bundgaard2020} conducted in Denmark reported that a mask recommendation did not have a statistically significant effect: in the study's primary composite outcome, 42 vs. 53 people tested positive in the intervention and control groups, respectively.   However, the vast majority of these positive results were from antibody tests, and given the antibody test's comparable incidence and false positive rates (approximately $2 \%$ and $0.8 \%$, respectively), a substantial fraction of the positive antibody tests in both the control and intervention groups are likely to be false positives, which would affect both the study's power and its statistical analyses~\cite{taleb_danish}.    Further false positives could arise from individuals who were infected before the study but for whom seroconversion did not occur until partway through the study.  However, false positives were not accounted for in the study's statistical analysis or conclusions.  If only the more reliable PCR tests are considered, then the reduction in infection due to masks (0 vs. 5 infections) is statistically significant ($p<0.05$).

\section*{Nonlinear effects}

We now describe a framework to account for the nonlinear aspects  of mask effectiveness.  Given that there is a threshold for the viral dose (the amount of the virus inhaled) below which the probability of infection is very small due to the innate immune system~\cite{yezli2011,nyt}, and given that the probability of infection $p$ will converge to one (for susceptible individuals) as the viral dose $v$ is increased without limit, the probability of infection as a function of viral dose $p(v)$ is described by a sigmoid function or S-curve (\cref{fig:convexity}, see Appendix for details).  (Concave curves have also been used to model dose response curves, but such an approach ignores threshold effects~\cite{sze2010, teunis2010}.)  For a single exposure event, we can define the dimensionless \textit{effective exposure} $\tilde v\equiv -\ln(1-p(v))$ such that the probability of infection is $1-e^{-\tilde v}$. Conveniently, the effective exposure is additive for independent exposure events, i.e. the total probability of infection is given by $1-e^{-\tilde v_T}$ where the \textit{total effective exposure} $\tilde v_T$ is simply equal to the sum of the effective exposures for each exposure event. 

\begin{figure}
\centering
\includegraphics[scale=0.9, trim={0 0 0 0}, clip]{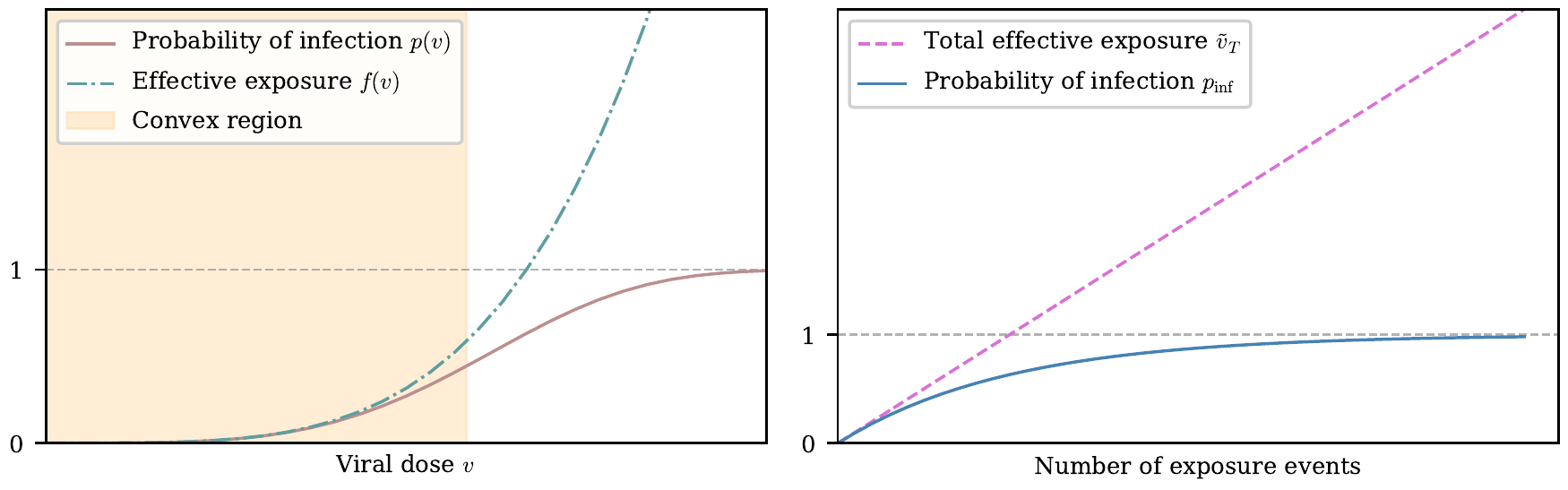}
\caption{\textbf{Left:} A representative function for a susceptible individual's probability of infection $p$ as a function of viral dose $v$ for a single exposure event, together with the effective exposure $\tilde v\equiv f(v)\equiv -\ln(1-p(v))$. $f(v)$ is convex for all $v$, while $p(v)$ is convex for sufficiently small $v$.  The convexity of $f(v)$ (which is demonstrated in the Appendix) yields an S-curve for $p(v)$.  Note that for any particular viral dose $v$, the effective exposure $\tilde v=f(v)$ can vary from individual to individual. 
\textbf{Right:} A depiction of how the total effective exposure $\tilde{v}_T$ and the probability of eventually becoming infected scale with the number of exposure events. The total effective exposure is the sum of the effective exposures from each exposure event; see Appendix for details.}
\label{fig:convexity}
\end{figure}

Because the probability of infection is a concave function of the total effective exposure (\cref{fig:convexity}), the protection afforded by a mask is super-linear in the percentage of exposures for which it is worn (e.g. wearing a mask twice as often is more than twice as effective; see \cref{fig:p_inf}).  These nonlinear effects can be substantial for high cumulative exposures.  Under such conditions, a mask may need to be worn for most or nearly all of the exposure events in order to provide significant protection; otherwise the individual is likely to be infected during the exposures for which the mask is not worn.
In the limit of an extremely high total exposure, a mask will of course not have an effect on the probability of infection since a susceptible individual will be infected with nearly 100\% probability regardless of whether or not the mask is worn.

 On the other hand, for low total exposures, the protection masks provide will be approximately proportional to the fraction of exposures for which they are worn.  
 It should be noted that the total exposure of individuals can vary within any given study, such that even if the overall probability of infection is low, most of those who were infected may have been subjected to high cumulative exposures. Studies with low overall probabilities of infection also have an additional difficulty, which is that large sample sizes will be necessary in order that there may be enough infections in the non-mask group to produce a statistically meaningful comparison.  In other words, for sufficiently low total exposure, the probability of infection will be quite low even without a mask, and so further reductions to this probability, even if proportionally large, will be small in absolute terms.

\begin{figure}
\centering
\includegraphics[scale=0.9, trim={0 0 0 0}, clip]{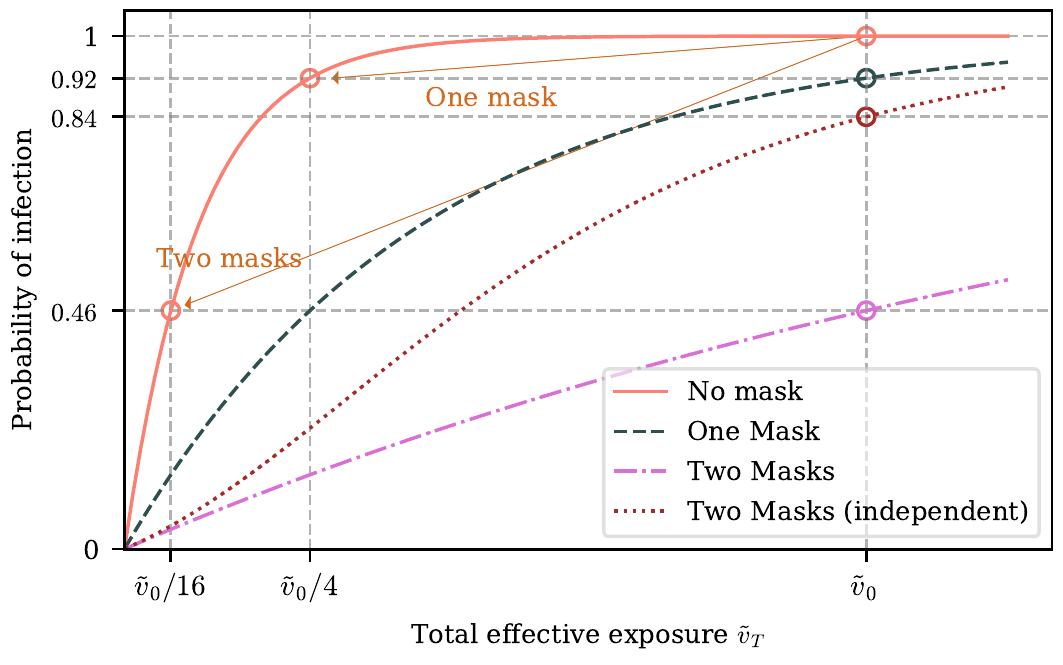}
\caption{The effect of both the susceptible and infected individual wearing a mask can be much larger than the effect of only one of them wearing a mask.  In the depicted example, the total effective exposure $\tilde{v}_0$ if neither the infected nor susceptible individual are wearing masks is such that the probability of infection $p_{\text{inf}}(\tilde{v}_0)$ is very close to 1.  If each mask reduces the effective exposure by a factor of $4$, then the probability of infection if only one of the two individuals is wearing a mask is $p_{\text{inf}}(\tilde{v}_0/4)=0.92$, i.e. a reduction in risk by a factor of $1.08$.   If both individuals are wearing a mask, however, the probability of infection is $p_{\text{inf}}(\tilde{v}_0/16)=0.46$, corresponding to a reduction in risk by a factor of $2.17$, which is greater than the product of the effects of each mask individually (shown by the red dotted curve).  
For illustrative purposes we have assumed that the infectious individual wearing a mask has the same effect as the susceptible individual wearing a mask, but relaxing this assumption will not qualitatively change the results; see Appendix for details.}
\label{fig:compound}
\end{figure}

We can also analyze certain compound effects that are not considered in most empirical studies. For instance, masks worn on both the infected and susceptible individuals may prevent a transmission event even if neither mask individually would have.  Furthermore, this compound effect may be super-linear: if the effect of only an infected individual wearing a mask is to reduce the infection probability by a factor of $p_1$ and the effect of only a susceptible individual wearing a mask is to reduce the infection probability by a factor of $p_2$, both individuals wearing a mask could reduce the infection probability by far greater than a factor of $p_1p_2$, especially for large total effective exposures. In the example shown in \cref{fig:compound}, the probability of transmission is reduced by only a factor of $1.08$ (a 7\% reduction) due to one or the other individuals wearing a mask, while if both wear a mask, the probability of transmission will be reduced by a factor of $2.17$ (a 54\% reduction).  Similarly, just as there can be a super-linear compound effect from both individuals wearing masks, there can also be super-linear compound effects when mask-wearing is combined with other behaviors that reduce exposure, such as social distancing. Nonlinear effects continue to accumulate when multiple individuals perform multiple behavioral changes that reduce exposure.  Recognizing these nonlinear effects is key to appreciating the effectiveness of transmission prevention policies. 

It should also be noted that the proportional risk reduction from masks is expected to be large because---due to the convexity of the S-curve (\cref{fig:convexity}) when exposure is low (as is likely for many exposure events)---the probability of the mask-wearer being infected is decreased by a greater factor than the decrease in the viral dose~\cite{Taleb} (see Appendix).   For these low exposure events, although the probability of infection may be small for any given potential transmission event, given multiple events, the large factors by which the probabilities of infection decrease due to this convexity can significantly reduce both the spread of the virus and the probability that the wearer eventually is infected.  In other words, wearing a mask may not only prevent the wearer from spreading viruses to others but may also have a surprisingly large protective effect for the mask-wearer.  Indeed, studies that analyze population-level data show that masks significantly reduce transmission~\cite{Howard2020, Lyu2020, Kenyon2020, Leffler2020, Cheng2020}.

In addition to the probability of infection, the implications of a nonlinear dose-response curve apply to several other outcomes as well.  In all of the above analyses, the probability of infection can be replaced with the probability of death or the probability of a particular degree of severity of symptoms, each of which can have a unique S-curve (that can also vary from individual to individual).  Thus, even when a mask does not prevent infection, it may reduce the severity of symptoms and the chance of long-term health damage or death.   It has been observed for the influenza virus that increasing the viral dose may lead to more adverse symptoms~\cite{influenza_dose, Virlogeux2015, Gandhi2020}, an effect that may also apply to SARS-CoV-2~\cite{Hamster,Gandhi2020,Cruise_mask,Food_plant, Bielecki2020}.

\section*{Conclusions}

Masks block some fraction of viral particles from dispersing from those who are infected and from infecting those who are susceptible and are understood to prevent disease transmission through this mechanism.  However, this simple understanding has been questioned based upon mixed empirical evidence.  Here we have shown that studies that did not find masks to be effective had limited statistical power and therefore do not imply that masks are ineffective.  The empirical evidence as a whole is thus consistent with current epidemiological recommendations to use masks during the COVID-19 pandemic. 

We note that psycho-social effects can reinforce the effect of prevention. The more individuals who wear masks, the less stigma that is associated with wearing them, which may make it more likely for others---including those who are infectious (whether symptomatic, pre-symptomatic, or asymptomatic)---to wear masks.  Masks may also serve as a reminder to take precautions that compound super-linearly with mask-wearing such as social distancing, although risk compensation, in which mask-wearing provides a sense of security that leads to higher risk taking, is also possible.  Effective public communication should emphasize that masks should be used in addition to and not as a substitute for other precautionary measures.

We have also shown that for many exposure events, masks will reduce the probability of infection by a greater factor than the factor by which they filter viral particles. This effect is also compounded non-linearly when both infected and susceptible individuals wear masks. When interpreted in light of this \textit{a priori} reasoning and the other considerations discussed above, the evidence indicates that, in addition to preventing the wearer from spreading respiratory infections, masks also protect the wearer from contracting them. The studies that did not find statistically significant effects prove only that masks cannot offer protection if they are not worn.

\subsection*{Acknowledgements}
This material is based upon work supported by the National Science Foundation Graduate Research Fellowship Program under Grant No. 1122374 and by the Hertz Foundation.  We thank Jeremy Rossman for helpful comments.

\appendix
\section*{Appendix}
\label{app:MA}

\subsection{Accounting for non-linearities in the effectiveness of masks}
In this section we develop a framework with which to understand the effect of masks.  We show that even if masks were to reduce the viral dose by only a modest factor, they may have a significantly larger impact on the probability of infection.   We  demonstrate that wearing a mask more frequently can super-linearly reduce one's chance of infection (e.g. wearing a mask 80\% of the time reduces one's probability of infection by more than twice as much as wearing a mask 40\% of the time).  We also show that when both infected and susceptible individuals wear masks, there can be a super-linear compound effect (e.g. if only infected individuals wearing masks reduces the probability of infecting susceptible individuals by a factor of 3 and only susceptible individuals wearing masks reduces the probability of being infected by a factor of 2, then if both wear masks the probability of infection will be reduced by a factor that is greater than $2\times 3=6$). 

\subsubsection{General Framework}
Although there is insufficient data to precisely describe the probability of infection as a function of the viral dose inhaled in a single exposure event, we can nonetheless derive some constraints on its shape. For a susceptible individual, the probability of infection (or any other outcome such as hospitalization or death) $p$ is a function of the viral dose $v$, i.e. the quantity of virus to which the individual is exposed. (This function $p(v)$ will vary from individual to individual based on biological factors, but should retain the general properties described below.)  For small $v$ the probability of a susceptible individual becoming infected will approach zero, and for large $v$ this probability will approach one, so $p(0) = 0$ and $p(\infty) = 1$. Since receiving two viral doses at once should not result in a lower probability of infection than the hypothetical in which the exposure to each viral dose could be modeled as an independent event, we have that  
\begin{equation}
\label{eq:p}
p(v_1 + v_2) \geq p(v_1) + p(v_2) - p(v_1)p(v_2)
\end{equation}
Equality will hold only in the absence of threshold effects; given that such effects are well established, we expect the inequality to be strict for small $v_1$ and $v_2$.  In order to characterize the set of functions satisfying \cref{eq:p}, we transform $p(v)$ using $p(v) \equiv 1-e^{-f(v)}$, or equivalently, $f(v)\equiv -\ln(1-p(v))$.  Eq.~\ref{eq:p} is then equivalent to 
\begin{equation}
f(v_1+v_2) \geq f(v_1) + f(v_2)
\end{equation}
Thus \cref{eq:p} is equivalent to $f(v)$ being convex. Choosing a convex $f(v)$ and then transforming back to $p(v)$ yields an S-curve (\cref{fig:convexity}), also known as a sigmoid function or sigmoid curve.  

When $f(v) \ll 1$, it can be shown by Taylor expansion that $p(v) \approx f(v)$. Thus, for small viral doses, $p(v)$ will be convex as well.  If a mask reduces the viral dose $v$ by a factor $b$~\cite{Booth2013, gawn2008}, then the mask will reduce the probability of infection (or of some other outcome denoted by $p(v)$ such as the probability of severe infection or death) by a factor of  $\frac{p(v)}{p({v/b})}$, which depends on $v$.  When $p(v)$ is convex, the factor by which the mask reduces the probability of infection will be greater than $b$ (since convexity implies that $p(v/b)<\frac{1}{b}p(v)+(1-\frac{1}{b})p(0)=\frac{1}{b}p(v)$).  Thus, for small exposures, masks can result in a surprisingly large reduction in the probability of infection.  We treat the impact of masks in more generality below, after introducing a framework for considering multiple exposure events.

The S-curve describes the probability of infection for a single exposure event. For $N$ independent exposure events, the probability of getting infected is $p_{\text{inf}} = 1 - \prod_{i=1}^N (1-p(v_i))$.  Using the form $p(v)=1-e^{-f(v)}$ as discussed above, 
\begin{equation}
    p_{\text{inf}} = 1 - e^{-\sum_{i=1}^Nf(v_i)}
\end{equation}
Defining the \textit{effective exposure} $\tilde v\equiv f(v)$ and defining $\tilde p(\tilde v)\equiv 1-e^{-\tilde v}$,
\begin{equation}
    p_{\text{inf}}=\tilde p (\tilde v_T)
\end{equation}
where $\tilde v_T=\sum_i \tilde v_i$ is the total effective exposure.  Considering the effective exposure $\tilde v$ rather than the actual dose $v$  is convenient since the effective exposure for repeated independent exposures is simply the sum of the individual effective exposures.  Note that for small effective exposures, the probability of being infected is approximately equal to the effective exposure, i.e. $\tilde p(\tilde v)\approx \tilde v$ for $\tilde v << 1$.  

\subsubsection{One mask}
Let $\gamma$ be the typical amount by which a mask reduces the effective exposure from a single exposure event---i.e. $\tilde v\rightarrow (1-\gamma)\tilde v$.  

Since $f(0)=0$ and $f(v)$ is convex, the simplest possible expression for $f(v)$ is the scale-free form $f(v)=(v/v_0)^\beta$, for some $v_0>0$ and $\beta>1$.  In this case, if a mask reduces the viral dose $v$ to $v/b$, $\gamma$ can be calculated exactly as $\gamma=1-b^{-\beta}$, regardless of $v$. For small exposures, the infection probability is roughly equal to the effective exposure, which is reduced by a factor greater than $b$ (i.e. $\frac{1}{1-\gamma}>b$) due to convexity ($\beta>1$), consistent with the analysis above.   This effect could potentially be quite large: e.g. for $\beta=4$, a mask filtering half of the viral particles ($b=2$) corresponds to a sixteen-fold reduction in effective exposure ($\gamma\approx 0.94$).
If we relax this assumption on the function $f$, then $\gamma$ becomes an effective parameter that may depend on the distribution of viral doses to which an individual is exposed.  Regardless of the precise form of $f(v)$, however, $\frac{1}{1-\gamma}>b$ will always hold due to the convexity of $f(v)$, i.e. masks will always have a disproportionately large effect on the effective exposure (and thus also on the infection probability when the effective exposure is small).  

Then, if a mask is worn for a fraction $\alpha$ of all exposures, the total effective exposure will be reduced from $\tilde v_T$ to $(1-\alpha\gamma)\tilde v_T$.   The probability of infection is thus
\begin{equation}
    \tilde p((1-\alpha\gamma)\tilde v_T)=1-e^{-(1-\alpha\gamma)\tilde v_T}
    \label{eq:onemask}
\end{equation}
(see \cref{fig:p_inf}).

Thus, we see that for any fixed $\gamma$ (mask effectiveness) and $\tilde v_T$ (total effective exposure without a mask), the benefit of wearing a mask is a convex function of the fraction $\alpha$ of the exposure events for which it is worn.  In other words, wearing a mask $x$ times as often will reduce the infection probability by more than a factor of $x$.  Thus, even if masks were 100\% effective ($\gamma=1$), a study in which participants wear masks 10\% of the time would need to have sufficient power to detect less than a 10\% reduction in the probability of infection. Our analysis therefore overestimates the true power of the studies.  

\subsubsection{Two masks}
To the extent that two masks together have an approximately linear effect on the effective exposure (e.g. if one person wearing a mask reduces effective exposure by $1-\gamma_1$ and the second person wearing a mask reduces effective exposure by $1-\gamma_2$, then both wearing masks reduces effective exposure by $1-\gamma_{12}\approx (1-\gamma_1)(1-\gamma_2)$), the effect on the probability of transmission will be super-linear, since the probability of infection $\tilde p(\tilde v)$ is concave in the effective exposure $\tilde v$.  In other words, especially for individuals who would have received a large total effective exposure without masks, both the susceptible and infectious individuals wearing masks will have a larger effect than would be calculated if each mask had an independent effect on the probability of transmission.  

If the effect of the two masks on the effective exposure is super-linear (i.e. $1-\gamma_{12}< (1-\gamma_1)(1-\gamma_2)$), then the effect on the probability of transmission will be super-linear to an even greater extent.  If the effect of the two masks on the effective exposure is sub-linear (i.e. $1-\gamma_{12} > (1-\gamma_1)(1-\gamma_2)$), then whether or not they still have a super-linear effect on the probability of transmission will depend on the total effective exposure.

(Note: Under the simplest possible form for $\tilde v=f(v)$, i.e. $f(v)=(v/v_0)^\beta$, if the mask on the infected individual reduces $v$ by a factor of $b_1$, the mask on the susceptible individual reduces $v$ by a factor of $b_2$, and together the masks reduce $v$ by a factor of $b_1b_2$, then the masks will have a linear effect on effective exposure, i.e. $1-\gamma_{12}=(1-\gamma_1)(1-\gamma_2)$.  Under other forms for $f(v)$ or assumptions about how the masks affect $v$, other behavior is possible.)

\begin{figure}
\centering
\includegraphics[trim = {0, 9cm, 0, 1.5cm}, scale = 0.85]{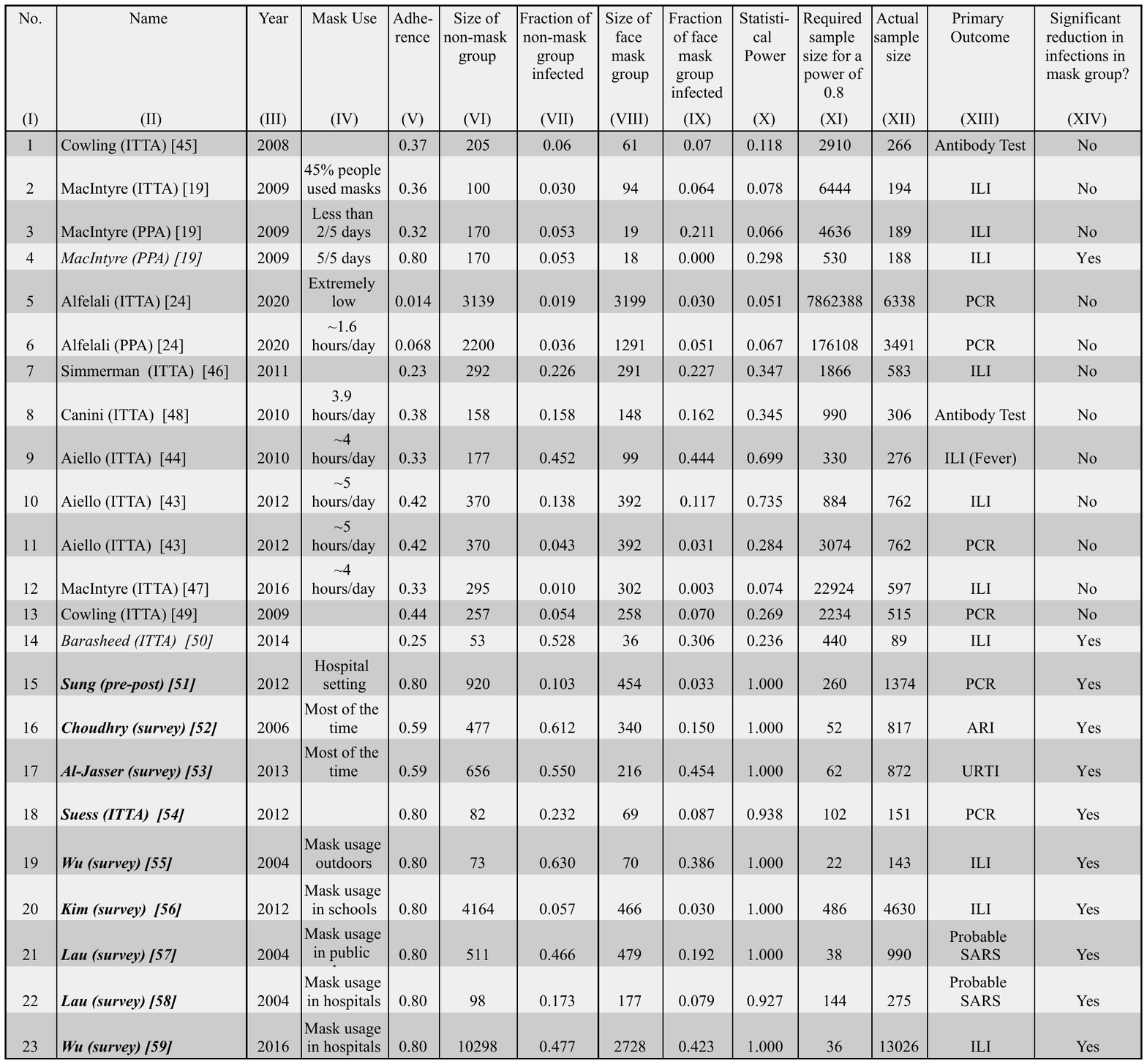}
\caption{{Summary of statistical power analysis.  Given the adherence levels reported in the studies the sample size necessary for a statistical power of 80\% for a two-tailed test and significance level of $0.05$ (assuming participants are equally divided between the non-mask and mask groups) is presented in column XI.   The statistical power given the actual sizes of the non-mask and mask groups is presented in column X.  These calculations were made for the case in which masks are 100\% effective; if masks are effective but not perfectly so, the necessary sample sizes for 80\% power (column XI) will be larger, while the statistical powers given the actual sample sizes (column X) will be lower.   Studies found to have greater than 80\% power are in bold (nos. 15-23), and studies that found a statistically significant reduction in infections in the mask group are italicized (nos. 4, 14-23).  Adherence is defined as the fraction of exposure for which masks were used; calculations of adherence for each study are presented in \cref{tab:adh}.  For studies that reported multiple analyses, each analysis is listed as its own entry (e.g. Aiello (2012)~\cite{Aiello2012} performed one analysis in which infection is defined by influenza-like illness (no. 10) and one analysis in which infection is defined by a positive PCR test result (no. 11)).  Note that only in the intention-to-treat analyses are participants randomly divided between the non-mask and mask groups; in survey and per-protocol analyses, which group a participant belongs to depends on whether or not that individual reported wearing a mask with a frequency above a threshold decided by the study.  \newline
\textbf{Abbreviations:} \textbf{ITTA}: Intention-to-treat analysis; \textbf{PPA}: Per-protocol analysis; \textbf{ILI}: Influenza-like illness; \textbf{ARI}: Acute respiratory infection; \textbf{URTI}: Upper respiratory tract infection; \textbf{PCR}: Polymerase chain reaction test (nasopharyngeal swab test); \textbf{SARS}: Severe Acute Respiratory Syndrome} }
\label{fig:power_table}
\end{figure}

\subsection{Power analyses}
Let $p_1$ and $p_2$ be the probabilities of getting infected in the non-mask (size $N_1$) and mask group (size $N_2$), respectively. Defining $\epsilon = p_1 - p_2$, the null hypothesis is $H_0 : \epsilon = 0$ and the alternate hypothesis is $H_1 : \epsilon \neq 0$. A test statistic is 
\begin{equation}
    W = \frac{ \hat{p_1} - \hat{p_2}}{\sqrt{ \hat p_1(1-\hat p_1)/N_1 + \hat p_2(1-\hat p_2)/N_2} } = \frac{\hat p_1 - \hat p_2}{\hat s} 
\end{equation}
where $\hat p_1$ and $\hat p_2$ refer to the observed fraction of infections, assumed to be normally distributed random variables whose means are $p_1$ and $p_2$ (this approximation is asymptotically exact).  We use the shorthand $\hat s$ for the denominator of $W$; note that $\hat s$ is an estimator for $s$, where $s^2=p_1(1- p_1)/N_1 + p_2(1- p_2)/N_2$ is the sum of the asymptotic variances of $\hat p_1$ and $\hat p_2$.  Asymptotically, $W-\epsilon/s$ follows a standard normal distribution.  Using the standard notation $\Phi(z_{1-\alpha/2}) = 1-\alpha/2$ where $\Phi(x)$ is the standard normal cumulative distribution function, the rejection region under $H_0$ for a significance level of $\alpha$ is given by the union of 
\begin{equation}
     W <- z_{1-\alpha/2} \text{ and } W > z_{1-\alpha/2}
\end{equation}
The various studies may use slightly different statistical tests, but the differences between tests should be small and will asymptotically disappear entirely.
  For any particular values of $\epsilon$ and $s$, the probability $W< -z_{1-\alpha/2}$ is asymptotically given by $\Phi(-z_{1-\alpha/2}-\epsilon/s)$ and the probability $W> z_{1-\alpha/2}$ is asymptotically given by $1-\Phi(z_{1-\alpha/2}-\epsilon/s)=\Phi(- z_{1-\alpha/2}+\epsilon/s)$.  Thus, given $\epsilon$ and $s$, the power, denoted by $1-\beta$ and equal to the probability that the null hypothesis is rejected if it is indeed false, is asymptotically given by

\begin{equation}
1-\beta= \Phi(- z_{1-\alpha/2}-\epsilon/s ) + \Phi(- z_{1-\alpha/2}+\epsilon/s ) \label{eq:power}
\end{equation}

Under the assumptions that masks are fully effective ($\gamma=1$) and that the probability of infection $p_{\text{inf}}$ decreases linearly with the adherence, the effect of mask usage is
\begin{equation}
    p_{\text{inf}} \rightarrow p_{\text{inf}}(1-a) \\
\end{equation}
where the adherence $a$ is the average fraction of exposure events for which the masks were used (see section 1 of the Appendix; here we use $a$ instead of $\alpha$ for the adherence to avoid confusion with the significance level).  Thus, for an infection probability $p_{\text{inf}} = p_1$ in the non-mask group (size $N_1$), the infection probability in the mask group (size $N_2$) will be $p_2 = p_1(1-a)$.  Thus, by estimating $p_1$ and $a$ for each study, we can use \cref{eq:power} to find power of each study given the sizes of their non-mask and mask groups, as well as the sample size (i.e. total number of participants) that would have been required for 80\% power.  For the latter estimate, we assume a study design in which the participants are evenly divided between the non-mask and mask groups (i.e. $N_1+N_2=2N_1=2N_2$) and rounded up the necessary sample size to the nearest even integer. 

For certain studies, some participants in the non-mask group used masks as well. In this case, adherence in both the mask group and non-mask group must be considered. Under the assumption that probability of infection decreases linearly with effective adherence, the probability of infection in the non-mask group $p_1$ is related to the probability of infection without masks $p_0$ by $p_1 = p_0(1-\gamma a_1)$ where $a_1$ is the adherence in the non-mask group and $\gamma$ is mask effectiveness. Then the probability of infection in the mask group will be $p_2 = p_0(1-\gamma a_2)$ where $a_2$ is the adherence in the mask group. The net adherence $a$ is defined by $p_2=(1-\gamma a)p_1$, which yields $a=\frac{a_2-a_1}{1-\gamma a_1}$.  In our analyses we assume $\gamma=1$, which leads to an overestimate for the net adherence $a$ of
\begin{equation}
    a= \frac{a_2-a_1}{1-a_1} \label{eq:net_adh}
\end{equation}

We estimate $p_1$ using the observed fraction of infections in the non-mask group $\hat p_1$. 
To check the robustness of our conclusions, we did a sensitivity analysis and found that if $\hat p_1$ differs from $p_1$ by a standard deviation (i.e. if we increase our estimate of $p_1$ by $\sqrt{\frac{1}{4N_1}}$), all studies that were under-powered ($<80\%$), except for one \cite{Aiello2012} remain under-powered. (To ensure robustness we used $\sqrt{\frac{1}{4N_1}}$ as the standard deviation, which is the maximum possible value of the true standard deviation $\sqrt{p_1(1- p_1)/N_1}$.)  If $\hat p_1$ underestimates $p_1$ by two standard deviations, another study~\cite{Aiello2010} would have greater than 80\% power under our assumptions.  It should be noted, however, that these assumptions overestimate the power in multiple ways (fully effective masks, overestimated adherence values, assuming a linear relationship between adherence and effectiveness, and the fact that individuals whose infections were not detected until after the start of the study could have actually been infected before they start of the study, i.e. before the mask intervention was implemented).

A more significant limitation of our analysis is in the difficulty in estimating adherence.  Adherence is often reported qualitatively, and even when quantitative, it is reported as the amount of time for which one wears a mask, which may differ from the fraction of exposures for which masks were worn.  To account for this difficulty, our strategy has been to consistently overestimate statistical power; to this end, we have erred on the side of overestimating adherence (see \cref{tab:adh}), and have also used other overestimating assumptions described in the previous paragraph.\newline\newline

\begin{longtable}{|l|p{1.5cm}|l|p{2cm}|p{13cm}|}
\caption {Adherence calculations for each study.} \label{tab:adh} 
\\
\hline
\scriptsize No. & \scriptsize Name & \scriptsize Year & \scriptsize Masks used by & \scriptsize Description and calculation \\ \hline

\scriptsize 1 & \scriptsize Cowling (ITTA) \cite{Cowling2008} & \scriptsize 2008 & \scriptsize Infected patients and their contacts & \scriptsize  Household study: 45\% of $21$ index cases used masks and 21\% of $61$ contacts wore masks. To overestimate adherence, we assume no transmission occurs while either the index patient or contact is wearing a mask. Neglecting correlations between whether or not the index patient wore a mask and the number of contacts of that index patient, an upper bound for the probability that either a contact or the index patient corresponding to that contact used a mask is 45\%+21\% = 66\% (this is likely an overestimate since households in which index patients wear masks and households in which contacts wear masks are almost certainly not mutually exclusive). In the control group, 30 \% of index patients and 1 \% of contacts used masks. Those classified as using masks used them often or always; therefore we assume that they used masks for 80\% of all exposures, a likely overestimate since the participants were asked to use masks only when they are not sleeping or eating. Therefore, the adherence in the mask group is estimated as  $0.66 \cross 0.8 = 0.53$, and adherence in the control group is estimated as $0.31 \cross 0.8 = 0.25$. This leads to a net adherence of 0.37 according to \cref{eq:net_adh}. 
\\ \hline

\scriptsize 2 & \scriptsize MacIntyre (ITTA) \cite{MacIntyre2009}& \scriptsize 2009 & \scriptsize Contacts of infected patients & \scriptsize  Household study over 5 days: Contacts were told to use masks when in the same room as the index patient. We consider only the surgical mask group (the other group was using P2 masks). On day 3, maximum adherence was reported: 45\% of contacts used masks for most of the time. We assume that those who used masks used them for 80\% of exposures, a likely overestimate since contacts did not use masks while sleeping, even if the child (index patient) was next to them in bed, and because the contacts could have been infected even if they were not in the same room as the index patient. The adherence is estimated as $0.45\cross0.8 = 0.36$ \\ \hline

\scriptsize 3 & \scriptsize MacIntyre (PPA)  \cite{MacIntyre2009}& \scriptsize 2009 & \scriptsize Contacts of infected patients & \scriptsize  Household study over 5 days (see row no. 2): Participants in this arm of the per-protocol analysis used masks for $<2$ out of 5 days. Overestimating adherence at 0.8  for 2 days gives adherence $ = 2/5 \cross 0.8 = 0.32$. \\ \hline

\scriptsize 4 & \scriptsize MacIntyre (PPA) \cite{MacIntyre2009}& \scriptsize 2009 & \scriptsize Contacts of infected patients  & \scriptsize Household study over 5 days (see row no. 2): Participants in this arm of the per-protocol analysis used masks for all 5 days. Overestimating adherence at 0.8  for 5 days gives adherence an adherence of $5/5 \cross 0.8 = 0.8$.
\\ \hline

\scriptsize5 & \scriptsize Alfelali (ITTA) \cite{Alfelali2020}& \scriptsize 2020 & \scriptsize Susceptible individuals  & \scriptsize Hajj study: From figure 2 of the study, we can only obtain approximate numbers since numerical data is not available in the figure. An average across the four days gives us the percentage of people using masks for various amounts of time. Using the upper bounds of the reported time ranges, we compute the average mask usage duration.  For the last time range (greater than 3 hours), we assume that masks were used on average for 5 hours. This leads to an average mask use of 0.778 hours and an adherence in the mask group of $0.778/24 = 0.032$. Participants in the control group used masks for 0.438 hours on average, yielding an adherence in the control group of $0.438/24 = 0.018$. The net adherence value is thus $0.014$ (\cref{eq:net_adh}). Note that the systematic review~\cite{Brainard2020} uses an older pre-print version of this study. \\ \hline

\scriptsize 6 & \scriptsize Alfelali (PPA) \cite{Alfelali2020} & \scriptsize 2020 & \scriptsize Susceptible individuals  & \scriptsize  Hajj study (see row no. 5): Those who wore masks were compared to those who did not.  The average mask use among those who wore masks was 1.637 hours; thus adherence = $1.637/24 = 0.0682$. \\ \hline

\scriptsize 7 & \scriptsize Simmerman (ITTA) \cite{Simmerman2011}& \scriptsize 2011 & \scriptsize Infected patients and their contacts  & \scriptsize Household study: We compare the hand-hygiene group with the hand-hygiene + mask group.  Only median (and not mean) mask usage was reported for the index and contact individuals; we therefore approximate the mean with the median. The median mask usage for the index patient was 35 minutes. The mean of median mask usage for contacts---parents, siblings and other relations---was 107.9 minutes. We estimate that index patients and contacts were in contact for 10.4 hours per day using data from a similar study~\cite{MacIntyre2016} (row no. 12). Adherence is therefore estimated as $\frac{107.9+35}{60\cross 10.4} = 0.23$ (see row no. 1 for why the index and contact mask usages were added together), a likely overestimate, given that the majority of the households resided in small one-bedroom apartments and thus were likely in contact for significantly greater than 10.4 hours per day on average.  Furthermore, contacts could have been infected outside of their homes.  Also, it was reported that 17.6\% of individuals in the control group used masks, meaning it was likely that those in the hand-hygiene-only group did as well (which would further reduce the net adherence). \\ \hline

\scriptsize 8 & \scriptsize Canini (ITTA) \cite{Canini2010}& \scriptsize 2010 & \scriptsize Infected patients  & \scriptsize Household study: Average mask use was $3.9$ hours per day (from table 3 of the study). We estimate that index and contact patients were in contact for 10.4 hours per day using data from a similar study~\cite{MacIntyre2016} (row no. 12).  Adherence is therefore estimated as $\frac{3.9}{10.4} = 0.38$, a likely overestimate given that contacts could have been infected outside their homes, or in their homes while not in contact with the index patient.  \\ \hline

\scriptsize 9 & \scriptsize Aiello (ITTA) \cite{Aiello2010}& \scriptsize 2010 & \scriptsize Susceptible individuals  & \scriptsize University residence hall: Mask usage was recorded inside the residence hall and they were used for $3.92$ hours per day. Assuming that residents spent 12 hours outside the halls, we exclude it from the adherence calculation. Adherence $ = \frac{3.92}{24-12} = 0.33$, a likely overestimate because participants were only encouraged but not required to use masks outside the residence halls, where they may be infected. In addition, the participants had left the residence halls for spring break, during which they were not required to wear masks.\\ \hline

\scriptsize 10 & \scriptsize Aiello (ITTA) \cite{Aiello2012}& \scriptsize 2012 & \scriptsize Susceptible individuals  & \scriptsize University residence hall: Masks were used for $5.08$ hours per day.  Adherence $ = \frac{5.08}{24-12} = 0.42$ (see row no. 9). \\ \hline

\scriptsize 11 & \scriptsize Aiello (ITTA) \cite{Aiello2012}& \scriptsize 2012 & \scriptsize Susceptible individuals  & \scriptsize University residence hall: Masks were used for $5.08$ hours per day. Adherence $ = \frac{5.08}{24-12} = 0.42$ (see row no. 9). \\ \hline

\scriptsize 12 & \scriptsize MacIntyre (ITTA) \cite{MacIntyre2016}& \scriptsize 2016 & \scriptsize Infected patients  & \scriptsize Household study: In the mask group, index patients were in contact with contacts for an average of 10.4 hours, and used masks for an average of 4.4 hours. The adherence in the mask group is thus estimated as $\frac{4.4}{10.4} = 0.42$. In the control group, average mask usage was 1.4 hours; adherence in the control group is thus estimated as $\frac{1.4}{10.4} = 0.13$. Net adherence is thus $0.33$ (\cref{eq:net_adh}). \\ \hline

\scriptsize 13 & \scriptsize Cowling (ITTA) \cite{Cowling2009}& \scriptsize 2009 & \scriptsize Infected patients and their contacts  & \scriptsize Household study: We compare the hand-hygiene group with the hand-hygiene + mask group. In the hand-hygiene + mask group, 49\% of index cases and 26 \% of contacts used a mask often or always. We therefore calculate adherence in the hand-hygiene + mask group as $ (0.49+0.26) \cross 0.8 = 0.60$ (see row no. 1). In the hand-hygiene group, 5 \% of contacts and 31 \% of index cases used masks, which leads to an adherence $ = (0.31+0.05) \cross 0.8 = 0.29$ in the hand-hygiene group. Net adherence is thus $0.44$ (\cref{eq:net_adh}). 
\\ \hline

\scriptsize 14 & \scriptsize Barasheed (ITTA) \cite{Barasheed2014}& \scriptsize 2014 & \scriptsize Susceptible individuals  & \scriptsize  Hajj pilgramage: 36 people were in the face mask group: 8 people never used a mask; 11 people used masks for $<4$ hours; 8 people used masks used for 5-8 hours; 9 people used masks for $>8$ hours (from table 2 of the study). Using the upper limits of the duration ranges (and 12 hours for the $>8$ hour group), adherence = $\frac{1}{36}(8\cross 0/24 + 11\cross4/24 + 8\cross 8/24 + 9\cross 12/24) = 0.25$.\\ \hline

\scriptsize 15 & \scriptsize Sung (Pre-post) \cite{Sung2012}& \scriptsize 2012 & \scriptsize Potentially infected individuals  & \scriptsize Visitors had to use face masks when they visited patients in their rooms and the incidence of infections was recorded among the patients. Although adherence was not reported, it is reasonable to assume that adherence was high since the study was conducted in a hospital where doctors and health care workers would have ensured that protocols are followed; in addition the visitors were in contact with patients only for a limited duration. We therefore assume an adherence of $0.8$ \\ \hline

\scriptsize 16 & \scriptsize Choudhry (Survey) \cite{Choudhry2006} & \scriptsize 2006 & \scriptsize Susceptible individuals  & \scriptsize Survey study for Hajj pilgrims: We consider the group of male pilgrims who reported using masks most of the time, compared to a group who did not use masks.  We assume that masks were not used while sleeping or eating, and note that the pilgrims remain susceptible to infection during such activities since they slept in shared tents. Allotting 10 hours per day for sleeping and eating and other activities during which masks were not worn, we estimate the adherence as $14/24 = 0.59$. 
\\ \hline

\scriptsize 17 & \scriptsize Al-Jasser (Survey) \cite{AlJasser2012}& \scriptsize 2013 & \scriptsize Susceptible individuals  & \scriptsize  Survey study for Hajj pilgrims: We consider the group of male pilgrims who reported using masks most of the time, compared to a group who did not use masks, and therefore estimate adherence as 0.59 (see row no. 16).\\ \hline

\scriptsize 18 & \scriptsize Suess (ITTA) \cite{Suess2012}& \scriptsize 2012 & \scriptsize Infected patients and their contacts  & \scriptsize  Household study: From figures 2 and 3 in the study, the average mask usage (across 8 days and across both seasons) among contacts and index patients is 69.4\% and 56.4\%, respectively.  From this data it is not impossible that in every household either the index patient or contacts were wearing masks; using this potential overestimate, we calculate adherence as $1 \cross 0.8 = 0.8$ (see row no. 1). \\ \hline

\scriptsize 19 & \scriptsize Wu (survey) \cite{Wu2004} & \scriptsize 2004 & \scriptsize Susceptible individuals  & \scriptsize Survey study: Face mask usage was reported only outside the home. Adherence was reported subjectively -- `Never', `Sometimes', `Always' (table 1 of the study). We compare the groups which used masks always and never used masks, and use an adherence value of $0.8$ for the `Always' group, a likely overestimate since participants could have been infected from household contacts. \\ \hline

\scriptsize 20 & \scriptsize Kim  (survey)\cite{Kim2011} & \scriptsize 2011 & \scriptsize Susceptible individuals  & \scriptsize Survey study among school children for influenza: Mask usage during school hours was reported as `continuous', `irregular', `not used'. We assume an adherence of 0.8 for the `continuous' group (and compare the infection rate to the group that did not use masks), a likely overestimate since children could be infected outside of school hours.\\ \hline

\scriptsize 21 & \scriptsize Lau (survey) \cite{Lau2004} & \scriptsize 2004 & \scriptsize Susceptible individuals  & \scriptsize Survey study during SARS epidemic: Mask usage was recorded only for public places. The study considered the frequent use of masks as using a mask, and occasional/seldom/no use was considered as not using a mask. We assume a value of 0.8 for adherence, a likely overestimate since people could have gotten infected at home where mask usage was not recorded and since some mask usage was possible in the non-mask group.
\\ \hline

\scriptsize 22 & \scriptsize Lau (survey) \cite{Lau2004_b} & \scriptsize 2004 & \scriptsize Susceptible individuals  & \scriptsize Survey study during SARS epidemic: Mask usage was recorded only during hospital visits to patients with SARS.  We use an adherence value of 0.8 for hospital settings (see row no. 15). For this study 0.8 is likely an overestimate since SARS infection could have occurred outside of the hospital as well.
\\ \hline

\scriptsize 23 & \scriptsize Wu (survey) \cite{Wu2016} & \scriptsize 2016 & \scriptsize Susceptible individuals  & \scriptsize Survey study for influenza-like illness.  Mask usage was recorded only during hospital visits.  We use an adherence value of 0.8 for hospital settings (see row no. 15). For this study 0.8 is likely a substantial overestimate since infection could have occurred outside of the hospital as well.\\ 
\hline

\end{longtable}
\clearpage
\begin{table*}[h!]
\scriptsize{
\caption {Studies not included in power analysis.} \label{tab:exc} 
\begin{tabular}{|l|l|p{13cm}|}
\hline
\multicolumn{1}{|c|}{Name} & \multicolumn{1}{c|}{Year} & \multicolumn{1}{c|}{Reason for exclusion from power analysis} \\ \hline
Shin \cite{Shin2018} & 2018 & Study was randomized for testing a common cold drug rather than mask usage, and mask usage was comparable in both of the groups. \\ \hline
Zhang \cite{Zhang2013}& 2013 &  Unknown adherence and incomplete data. \\ \hline
Jolie \cite{Jolie1998}& 1998 & Animal to human transmission: We consider only human to human transmission for our analysis. \\ \hline
Tahir \cite{Tahir2019}& 2019 & Animal to human transmission: We consider only human to human transmission for our analysis. \\ \hline
Larson \cite{Larson2010} & 2010 & Mask adherence was reported to be `poor' but neither the percentage of participants using masks nor the duration of mask usage was reported, so we could not make an estimate for the adherence. \\ \hline
Emamian \cite{Emamian2013} & 2013 & Survey study for Hajj pilgrims:  Adherence for mask usage was reported only as `Yes' or `No'. Even occasional use of mask was considered as `Yes'. Since adherence data stratified by frequency and/or duration was not reported, we could not make an estimate for the adherence. \\ \hline
Deris \cite{Deris2010} & 2010 & Survey study for Hajj pilgrims: Adherence for mask usage was reported only as `Yes' or `No'. Since adherence data stratified by frequency and/or duration was not reported, we could not make an estimate for the adherence.  \\ \hline
Uchida \cite{Uchida2017} & 2017 & Survey study for children. Mask usage was reported as `using masks at any time or place'. Since adherence data stratified by frequency and/or duration was not reported, we could not make an estimate for the adherence. \\ \hline
Balaban \cite{Balaban2012} & 2012 & Survey study for Hajj pilgrims: Adherence for mask usage was reported only as `Yes' or `No'. Since adherence data stratified by frequency and/or duration was not reported, we could not make an estimate for the adherence. \\ \hline
Zein & 2002 & Study not available.\\ \hline

\end{tabular}
}
\end{table*}

\bibliography{references}{}
\bibliographystyle{naturemag}

\end{document}